# Superconductivity of the FeSe/SrTiO$_3$ Interface in the View of BCS-BEC Crossover

Shuyuan Zhang[1,2], Guangyao Miao[1,2], Jiaqi Guan[1,2], Xiaofeng Xu[1,2], Bing Liu[1,2], Fang Yang[1], Weihua Wang[1], Xuetao Zhu[1,2,3*], Jiandong Guo[1,2,3,4*]

[1] *Beijing National Laboratory for Condensed Matter Physics and Institute of Physics, Chinese Academy of Sciences, Beijing 100190, China*

[2] *School of Physical Sciences, University of Chinese Academy of Sciences, Beijing 100049, China*

[3] *Songshan Lake Materials Laboratory, Dongguan, Guangdong 523808, China*

[4] *Beijing Academy of Quantum Information Sciences, West Bld.#3, No.10 Xibeiwang East Rd., Haidian District, Beijing 100193, China*

**Abstract:**

**In paired Fermi systems, strong many-body effects exhibit in the crossover regime between the Bardeen-Cooper-Schrieffer (BCS) and the Bose-Einstein condensation (BEC) limits. The concept of the BCS-BEC crossover, which is studied intensively in the research field of cold atoms, has been extended to condensed matters. Here, by analyzing the typical superconductors within the BCS-BEC phase diagram, we find that FeSe-based superconductors are prone to shift their positions in the BCS-BEC crossover regime by charge doping or substrate substitution, since their Fermi energies and the superconducting gap sizes are comparable. Especially at the interface of a single-layer FeSe on SrTiO$_3$ substrate, the superconductivity is relocated closer to the crossover unitary than other doped FeSe-based materials, indicating that the pairing interaction is effectively modulated. We further show that hole-doping can drive the interfacial system into the phase with possible pre-paired electrons, demonstrating its flexible tunability within the BCS-BEC crossover regime.**

Based on the assumption of Fermi systems with attractive interaction, Bardeen-Cooper-Schrieffer (BCS) theory [1,2] has been well applied not only to electronic systems, but also applicable to other Fermi systems, such as the superfluid state of paired Fermionic atoms [3]. Accordingly the BCS picture can be linked to the concept of Bose-Einstein condensation (BEC), as demonstrated in experiments by directly tuning the attractive interaction or scattering length of ultra-cold atoms [4-7]. In this context, the phase diagram of a paired Fermi system is depicted by a BCS state of Fermions at one limit and a BEC state of bosonic molecules at the other limit, and scaled by $k_F\xi_{pair}$, where $k_F$ is the Fermi momentum, and $\xi_{pair}$ the pair coherence length or pair size, as shown in Fig.1 with $k_F\xi_{pair}$ v.s $T_C/T_F$ as the parameters ($T_C$: the critical temperature, and $T_F$: the Fermi temperature). In the BCS limit, overlapping Cooper pairs with large $\xi_{pair}$ are formed and condensed at $T_C$, accompanied with a short inter-particle distance, $1/k_F$, due to the large Fermi energy ($E_F$), resulting in $k_F\xi_{pair} \gg 1$. In the BEC limit, tightly bound molecule-like electron pairs have short $\xi_{pair}$ in real space, accompanied with large $1/k_F$ in the dilute molecule gas, i.e., $k_F\xi_{pair} \ll 1$. Those bosonic molecules, i.e., Bose liquid, condense into the superfluid state below $T_C$. Between these two limits, incoherent Fermion pairs are preformed due to the Fermi surface instability at the temperature $T_{pair}$ (higher than $T_C$), leading to the formation of the so-called pseudogap phase. Note that, considering the experimental accessibility, $\Delta/E_F$ ($\Delta$: the gap associated with the paired Fermions) v.s. $T_C/T_F$ can be introduced as the parameters of the BCS-BEC diagram of superconductors, since there is the monotonic one-to-one mapping between $\Delta/E_F$ and $k_F\xi_{pair}$, as shown in the Supplementary Materials (SM) [8].

There is a special point, called unitary point, defined by $k_F\xi_{pair} = 1$, where the strongest many-body effects may take place, and all thermodynamic and transport properties are not determined by the scattering length [4,5]. At the unitary point, $T_C/T_F \sim$ 0.167 [9] and $\Delta/E_F \sim 0.44$ [10] (as measured by $^6Li$ atom gas experimentally). The region near this point is the so-called BCS-BEC crossover regime, which is a fascinating subject to study the many body interactions [11,12]. Specifically, the analyses of the BCS-BEC crossover have been extended from cold atoms to superconductors for the understanding of the many-body effects involved in superconductivities. In 1990s, Uemura analyzed $T_C/T_F$ to investigate the BCS-BEC crossover behaviors of different types of superconductors [11,13-15], and found that cuprates possess a high value of $T_C/T_F \sim 0.05$, closer to the unitary point than other superconducting materials discovered thus far, including conventional superconductors [15] and heavy fermions systems [14-16]. In Fig. 2 (a), we list the data of representative superconducting cuprates in the scale of $T_C/T_F$. Although they exhibit increased values relative to the BCS and heavy fermion superconductors, they are still close to each other, apart from the unitary point.

To obtain superconductors close to the unitary point is essential to realize the investigations and further tuning of BCS-BEC crossover behaviors. The iron based superconductors provided new opportunities to this route. For instance, in FeSe, the low $E_F$ results in $\Delta/E_F = 0.3$ for the hole pocket (FeSe-h) and $0.83 - 1$ for the electron pocket (FeSe-e) [17,18], closer to the unitary point than cuprates [Fig. 2 (a)]. The potential BCS-BEC crossover behaviors were observed in $Fe_{1+y}Se_xTe_{1-x}$ [19,20], and the value of $\Delta/E_F$ can be tuned from 0.16 to 0.5 [21]. However the multiband electronic structure of iron pnictides [22] makes the investigations extremely complicated. And the crossover characters, such as the pseudogap and the giant fluctuation of diamagnetism, of FeSe are not observed [23,24]. In this study, we identify that the recently discovered superconducting interface of a single-layer FeSe on SrTiO$_3$ (STO) (1uc-FeSe/STO) [25,26] is an ideal platform for the exploration of BCS-BEC crossover behaviors: it exhibits the superconductivity with single electron pocket [27] located close to the BCS-BEC crossover unitary. By charge doping [28,29] and the introduction of interfacial interaction [30-33] by different substrates, it is possible to effectively tune the superconductivity in the BCS-BEC crossover regime. For a demonstration, the possible electron pre-pairing regime is approached by hole-doping to the 1uc-FeSe/STO.

Figure 2 (b) shows the BCS-BEC phase diagram of FeSe and FeSe-based superconductors scaled by $T_C/T_F$ v.s. $\Delta/E_F$. In the quasi-two-dimensional (quasi-2D) systems, like cuprates and Fe-based superconductors, although there are slight changes in the phase diagram compared to three dimensional continuum theory [34,35] due to the Berezinskii-Kosterlitz-Thouless transition and the discrete lattice, the phase diagram discussion based on the Fermi momentum and scattering length is still valid. For the bulk FeSe ($T_C$ ~ 8 K), two bands contribute to the Fermi surface: a hole band (FeSe-h) at Γ point with the Fermi energy $E_F^h$ ~ 10 meV and an electron band (FeSe-e) at M point with the Fermi energy $E_F^e$ ~ 3 meV, with the values of $\Delta/E_F$ ~ 0.3 for FeSe-h and 0.83 – 1 for FeSe-e, and $T_C/T_F$ ~0.07 for FeSe-h and ~0.23 for FeSe-e [17,18]. The superconductivity from either band is away from the BCS limit. The superconductivity originated from FeSe-e is even over the unitary point towards the BEC limit. With electron doping, e.g., in (LiFe)OHFeSe [36], $E_F$ increases to 43 meV from ~ 3 meV of FeSe (only FeSe-e is relevant since FeSe-h does not contribute to the Fermi surface any more), and the superconductivity shifts towards the BCS limit with $\Delta/E_F$ decreasing to ~ 0.24 and $T_C/T_F$ to ~ 0.08. For K$_{0.8}$FeSe [37], $E_F$ increases further to 60 meV, and $\Delta/E_F$ decreases to ~ 0.19 and $T_C/T_F$ decreases to ~ 0.05. It is clearly indicated that FeSe is a rare system that the superconductivity can be tuned in a broad range of the BCS-BEC crossover regime.

With the interface introduced, the 1uc-FeSe/STO [25], 1uc-FeSe/TiO$_2$ [38,39] or 1uc-FeSe/BaTiO$_3$ [40], exhibits larger value of $T_C/T_F$ and $\Delta/E_F$ than other superconductors as shown in Fig. 2. Taking the 1uc-FeSe/STO as the example, with $T_C$ ~ 65 K, $\Delta$ ~20 meV, $E_F$ ~ 56 meV [27,30,33,41], $\Delta/E_F$ increases to ~ 0.36 and $T_C/T_F$ to 0.1, getting close to the unitary point ($\Delta/E_F$ ~ 0.44 [10], $T_C/T_F$ ~ 0.167 [9]). The non-Fermi liquid behaviors are expected and indeed have been identified by the spectrum continuum in the high energy range of band structures observed by angle-resolved photoemission spectroscopy measurements [27,30]. It has been reported that both the electron transfer from STO to FeSe [27] and the non-adiabatic interfacial electron-phonon interaction (EPI) induced by the optical phonon of STO with the energy of 97 meV are responsible for the superconductivity enhancement [30,31,33].

These interfacial systems can be tuned near the unitary point, making them ideal platforms for the study of the BCS-BEC crossover behaviors. For example, if tuned further toward the BEC limit by hole doping to reduce the Fermi energy, the interfacial FeSe/Oxides system can possibly enter the pre-pairing region. Here, by depositing 7,7,8,8-tetracyanoquinodimethane (TCNQ) molecules on the surface of 1uc-FeSe/STO we dope holes to FeSe. The details of sample preparation and characterization are described in the SM [8]. Figure 3 (a) and (b) show the STM images of TCNQ molecules on the surface of 1uc-FeSe/STO. With the hole doping by TCNQ molecules, the electron density loss in FeSe is 0.05 e$^-$ per unit cell [42] at the position near point #5 in Fig. 3 (a). The corresponding $E_F$ is reduced from 56 meV on the pristine surface (at point #1) to 37 meV adjacent to the molecule (at point #5), resulting in the suppression of superconductivity at the experiment temperature (4.9 K) [42]. In Fig. 3 (c), at point #1 with a distance about 11 unit cells away from the TCNQ molecules, the STS shows a U-shaped superconducting gap with two obvious coherence peaks, which exhibits the same characteristics as on the pristine superconducting surface. As the position laterally approaches the TCNQ molecules from #1 to #5, the coherent peak disappears gradually without changing the gap size, as shown in Fig. 3 (c). Considering the lowered $E_F$, the hole-doped system is shifted toward the side of BEC regime in the phase diagram. The observed unchanged gap with suppressed coherent peak may be related to the spectral signature of the electron pre-pairing. More solid experimental evidence of the pseudogap state, especially the *in situ* $T_C$ measurement currently unavailable for the molecular absorbed samples, is needed in the future studies.

Besides the interface electron transfer from STO to FeSe that has been revealed to be responsible for the superconductivity modification [27], the bosonic mode introduced by the substrate is also indispensable [30,31,33]. It is intriguing to understand how the involved boson mode ($\hbar\omega_B$, the energy of the pairing glue) changes at the interface within the picture of BCS-BEC crossover. We summarize the $\hbar\omega_B/E_F$ values of coupled boson modes measured in experiments, and realize that there is a positive correlation between $\hbar\omega_B/E_F$ and the position in BCS-BEC phase diagram for FeSe-based superconductors. In the bulk FeSe, the magnetic excitation of ~ 4 mV has been observed by inelastic neutron scattering [43]. It is comparable to $E_F$, giving $\hbar\omega_B/E_F^h$ ~0.4 for FeSe-h band and $\hbar\omega_B/E_F^e$ ~1.3 for FeSe-e band. For (LiFe)OHFeSe and K$_{0.8}$FeSe, the magnetic excitation energies are 21 meV [44] and 14 meV [45], giving the values of $\hbar\omega_B/E_F$ ~ 0.49 and ~ 0.23, respectively. Coincident with the decrease of $\hbar\omega_B/E_F$ in FeSe-e, (LiFe)OHFeSe, FeSe-h and K$_{0.8}$FeSe, independent experiments indicate the same decreasing tendency of $\Delta/E_F$, as shown in Fig. 2(b). Although such a positive correlation is not dictated by any established theory, it possibly leads to new clues for the search of the pairing interaction in unconventional superconductors.

Now we turn to 1uc-FeSe/STO where the magnetic excitations of FeSe and the phonon modes of the oxide substrate are believed to contribute together to the superconductivity [46]. The possible magnetic excitations give $(\hbar\omega_B/E_F)_{spin}$ ~ 0.17 - 0.33 ($\hbar\omega_B$ ~ 9.6 - 18.2 meV) [47] and the optical phonon of the oxide substrate gives $(\hbar\omega_B/E_F)_{phonon}$ ~ 1.7 ($\hbar\omega_B$ ~ 97 meV) [30-33]. Considering the positions of the FeSe-derived superconductors in the BCS-BEC phase diagram, i.e., the FeSe-h, (LiFe)OHFeSe and $K_{0.8}$FeSe are relatively close to the BCS regime, the FeSe-e is relatively close to the BEC regime, while the 1uc-FeSe/STO is in the middle and close to the crossover unitary, the cooperation of the FeSe magnetic excitations and the STO phonons must be essential in the superconductivity at the interface. Following this route, to choose the substrate with the appropriate phonon energy at the FeSe/Oxides interfaces might be effective to tune the BCS-BEC crossover behaviors in one condensed matter system. Several candidate oxide substrates with suitable optical phonon energies are listed in Table I.

In summary, we show that the superconducting behaviors of FeSe and the derived superconductors can be understood within the BCS-BEC phase diagram. Especially, the superconductivity of FeSe/oxide interfaces locates closest to the BCS-BEC unitary point. At these interfaces, besides the modulation of the electron density due to the interfacial charge transfer, the substrate optical phonon can cooperate with the original pairing glue of FeSe to get multiple bosons involved in the interfacial superconductivity. This results in the effective tuning of the BCS-BEC crossover behaviors and therefore the significantly modified superconductivity. We demonstrate the flexible tunability by the hole doping to the 1uc-FeSe/STO that drives the system into the electron pre-pairing phase at 4.9 K.

**Acknowledgements**

The authors thank the discussions with Profs. E. W. Plummer and Jiandi Zhang. The work was supported by the National Key R&D Program of China (Nos. 2017YFA0303600, 2016YFA0300600 & 2016YFA0202300), the National Natural Science Foundation of China (No. 11634016), the Strategic Priority Research Program (B) of the Chinese Academy of Sciences (No. XDB07030100), and the Research Program of Beijing Academy of Quantum Information Sciences (No. Y18G09). X.Z. was partially supported by the Youth Innovation Promotion Association of Chinese Academy of Sciences (No. 2016008). W. W was partially supported by the Hundred Talents Program of the Chinese Academy of Sciences.

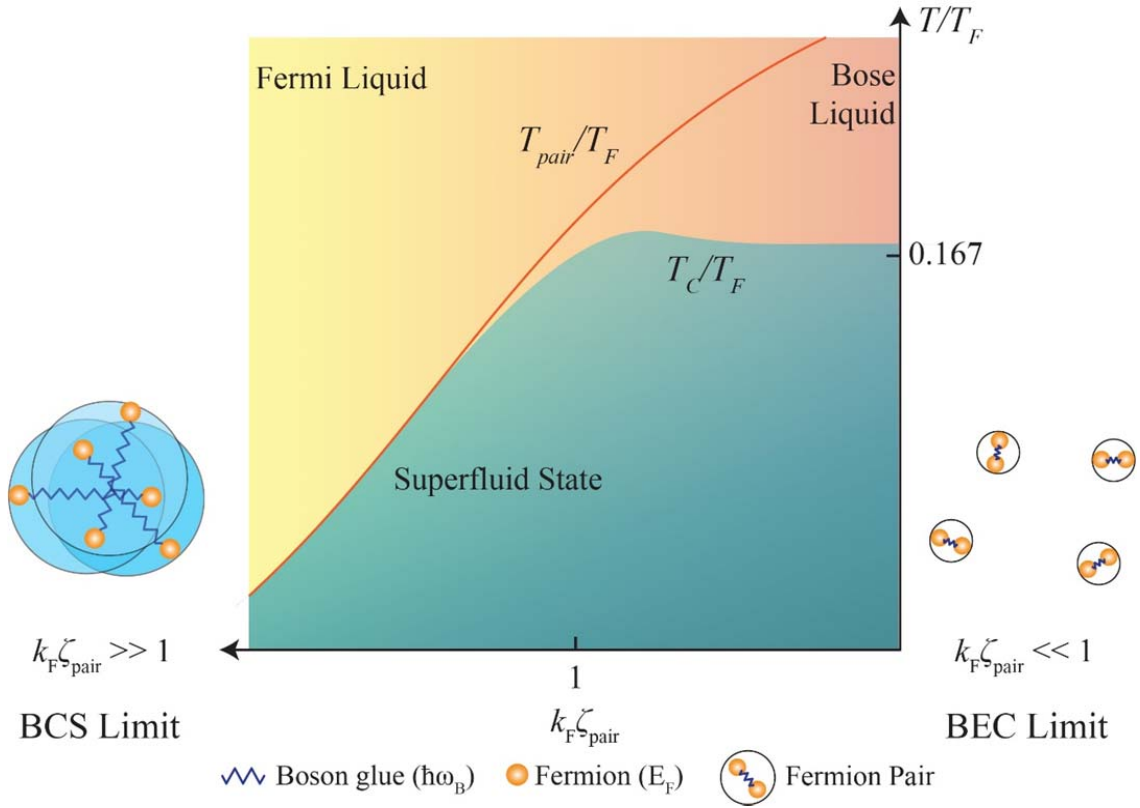

**Figure 1| The BCS-BEC phase diagram.** The quantitative schematic of the temperatures of electron pairing ($T_{pair}$) and superfluid state ($T_C$) as the functions of $k_F\xi_{pair}$. The relationship between parameters in the phase diagram was shown in the Supplementary Materials [8]. The region between the solid orange line $T_{pair}/T_F$ and the superfluid state $T_C/T_F$ represents the electron pre-pairing, *i.e.* pseudogap. The vertical axis is normalized to the Fermi temperature $T_F$. The left and right panels show the corresponding electron pairs in the BCS and the BEC limit, respectively.

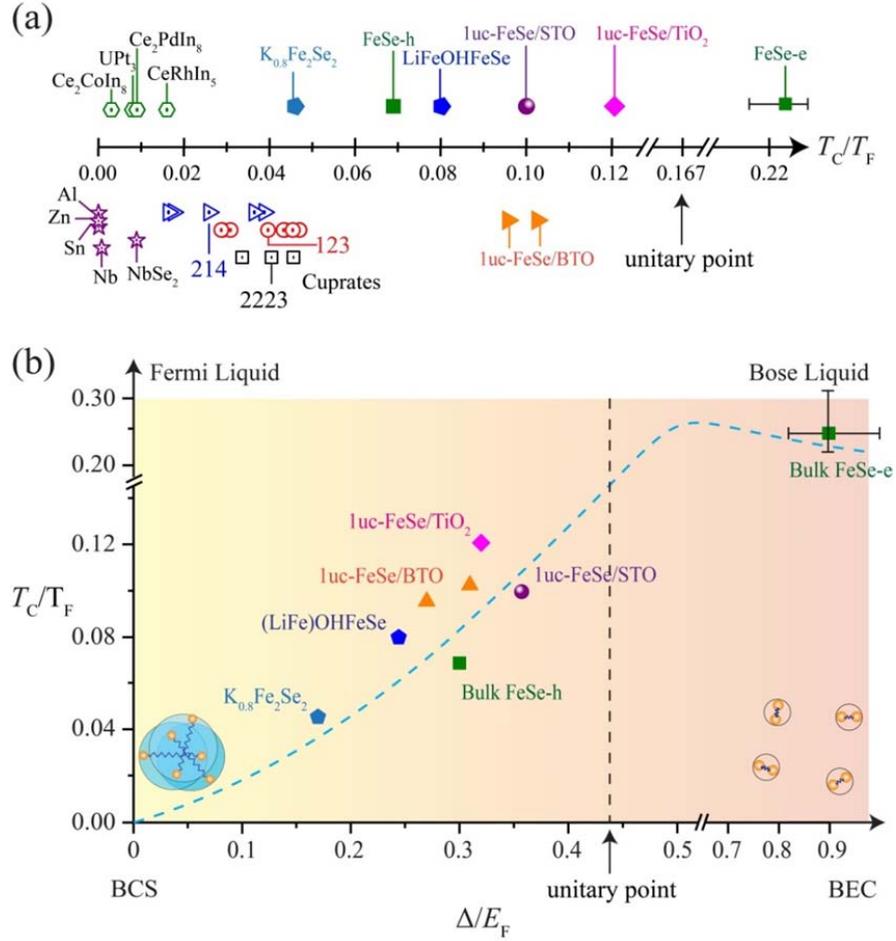

**Figure 2| BCS-BEC crossover analyses of different superconducting systems. (a)** The $T_C/T_F$ data of typical superconducting systems, including conventional superconductors (purple stars) [11,13,15], heavy fermion superconductors (green hexagons) [16], cuprates [blue triangles for $La_{2-x}Sr_xCuO_4$ (214), red circles for $YBa_2Cu_3O_y$ (123) and black squares for $Tl_2Ba_2Ca_2Cu_3O_{10}$, $Bi_{2-x}Pb_xSr_2Ca_2Cu_3O_{10}$, and $(Tl_{0.5}Pb_{0.5})Sr_2Ca_2Cu_3O_9$ (2223)] [11,13,15], and FeSe-based superconductors (the navy pentagon for $K_{0.8}Fe_2Se_2$ [37], blue pentagon for (LiFe)OHFeSe [36], green squares for FeSe bulks [17,18], purple circle for 1uc-FeSe/STO [33], pink square for 1uc-FeSe/TiO$_2$ [38], and orange triangle for 1uc-FeSe/BaTiO$_3$ [40]). The detailed parameters of these systems are listed in the Supplementary Materials [8]. The BCS-BEC unitary point is labeled for reference. **(b)** Illustration of the evolution of FeSe-based superconductors in the BCS-BEC crossover phase diagram scaled by $T_C/T_F$ v.s. $\Delta/E_F$. The data are taken from Ref. [17,18] for FeSe bulks, Ref. [36] for (LiFe)OHFeSe, Ref. [37] for $K_{0.8}Fe_2Se_2$, Ref. [40] for 1uc-FeSe/BTO, Ref. [38] for 1uc-FeSe/TiO$_2$, and Ref. [33] for 1uc-FeSe/STO. The dashed curve is the schematic drawing of $T_C/T_F$ as is in Fig. 1.

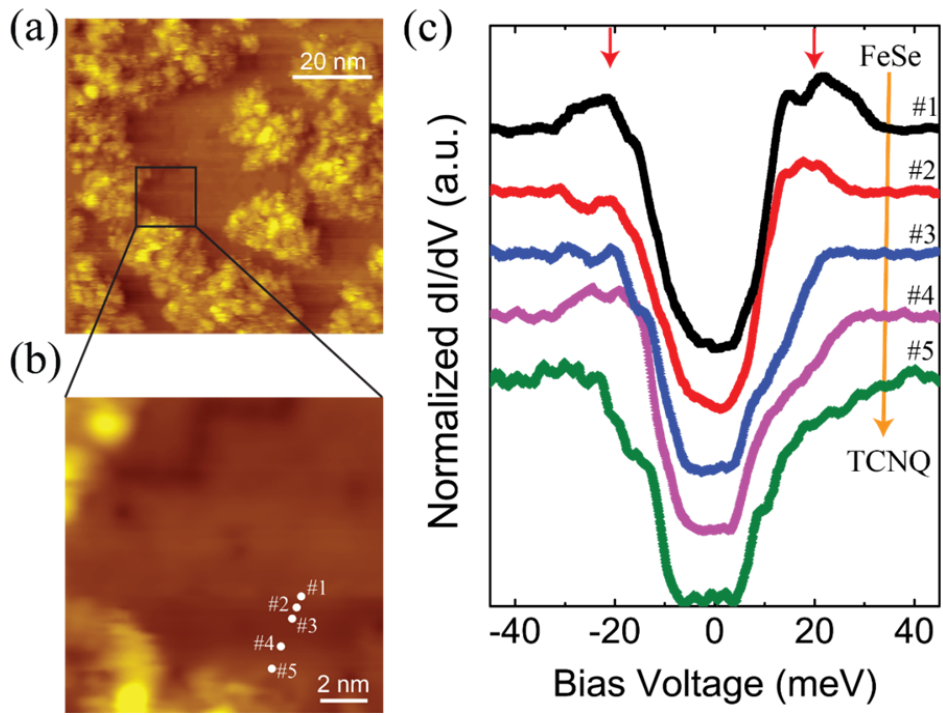

**Fig. 3| Observation of pre-pairing in hole-doped 1uc-FeSe/STO.** (**a**) The STM image (80 × 80 nm², 5.0 V/50 pA) of TCNQ molecules adsorbed on the surface of 1uc-FeSe/STO. (**b**) Zoom-in image (15 × 15 nm², 2.0 V/50 pA) of the area marked by the square in (**a**). (**c**) Normalized dI/dV spectra with tip laterally approaching the TCNQ molecules from position #1 to #5 as marked in (**b**).

Table I. Possible substrates which can be used to tune the BCS-BEC at the FeSe/Oxides interface. The $E_F$ of 1uc-FeSe/STO is taken as the reference.

| Energy Scale (meV) | Substrate | Phonon Energy (meV) |
|---|---|---|
| < $E_F$ | GaAs [48] | 34 |
| | RbF [49] | 35 |
| ~ $E_F$ | NiO [50] | 68 |
| | CaF$_2$ [48,49] | 56 |
| > $E_F$ | DyScO$_3$ [51] | 79 |
| | MgO [52] | 81 |
| | BaTiO$_3$ [32,53] | 88 |
| | SrTiO$_3$ [32] | 97 |
| | KTaO$_3$ [54] | 103 |
| | V$_2$O$_5$ [55,56] | 124 |
| | MoO$_3$ [55] | 124 |

# Supplementary Materials for

# Superconductivity of the FeSe/SrTiO$_3$ Interface in the View of BCS-BEC Crossover

The following Supplementary Materials include the descriptions of the basic concept and parameters of BCS-BEC crossover, the sample preparations and characterizations, as well as the BCS-BEC parameters of different superconducting systems.

I.   Parameters of the BCS-BEC phase diagram:

The phase diagram of interacting Fermi system as shown in Fig. 1 is scaled by a dimensionless parameter $1/(k_F a)$, where $a$ is the scattering length charactering the pairing interaction [5]. In the BCS limit, the weak attraction has a negative scattering length, and $1/(k_F a) \rightarrow -\infty$. In the BEC limit, $1/(k_F a) \rightarrow +\infty$. The unitary point is defined by $1/(k_F a) = 0$ with the scattering length divergence. The BCS-BEC crossover is the regime within $|1/(k_F a)| \lesssim 1$. And the BCS-regime and BEC-regime are characterized by $1/(k_F a) \lesssim -1$ and $1/(k_F a) \gtrsim 1$, respectively [57].

In ultra-cold atom experiments, the parameter $a$ can be tuned directly by the Feshbach resonance [6,7]. However, in condensed matter, the interaction is determined intrinsically and hard to be tuned. Thus in Fig. 1, we plot the phase diagram scaled by $k_F \xi_{pair}$ which is more apparent for condensed matter experiments. The relationship between $k_F \xi_{pair}$ [the horizontal axis in Fig.1(a)] and $1/(k_F a)$ was shown in Ref. [57]. For instance in the BCS limit, $k_F \xi_{pair} \sim e^{\pi/(2 k_F |a|)}$ [4]. Clearly when $1/(k_F a) = 0$, $k_F \xi_{pair} = 1$.

The value of $T_C/T_F$ and $\Delta/E_F$ are the functions of $k_F \xi_{pair}$ as schematically plotted in Fig. 1. As for the BCS-BEC crossover analyses of superconductors shown in Fig. 2, we use $\Delta/E_F$, instead of $k_F \xi_{pair}$ or $1/(k_F a)$, due to the lack of accurate parameter $k_F \xi_{pair}$ or $1/(k_Fa)$ in the published experimental data. In the BCS limit, $\Delta = \frac{8}{e^2}E_F e^{\frac{\pi}{2ak_F}}$; in the BEC limit, $\Delta = 4E_F/\sqrt{3\pi ak_F}$ [58]. There is a monotonic one-to-one mapping between the values of $\Delta/E_F$ and $1/(k_Fa)$. In the BCS limit, $k_BT_C = \frac{8E_F e^\gamma}{\pi e^2}e^{\frac{\pi}{2ak_F}}$, where $k_B$ is Boltzmann constant, $\gamma$ is the Euler constant with $\frac{e^\gamma}{\pi} = 0.567$ [57]. Then we have $\frac{2\Delta}{k_BT_C} = 3.5$ in the BCS limit which corresponds to the BCS theory [1].

## II. Sample Preparations and Characterizations

All the SrTiO$_3$ substrates used in this study were 0.5% Nb-doped SrTiO$_3$(001). The monolayer FeSe films were grown by the molecular beam epitaxy (MBE) method and were characterized by scanning tunneling microscopy. The 1uc-FeSe/STO samples were post-annealed at 470 °C for 6 hours in ultra-high vacuum to make the monolayer FeSe superconducting. The TCNQ molecules were evaporated onto the 1uc-FeSe/STO sample from a low-temperature evaporator at 390 K as described in Ref. [42].

The STM topographic images were acquired in constant-current mode with the bias voltage applied to the sample with respect to the tip. The STS were measured at 4.9 K with a bias modulation of 1 mV at 987.5 Hz. All the dI/dV spectra in the manuscript were normalized by a background defined by a polynomial function as shown in Fig. S1.

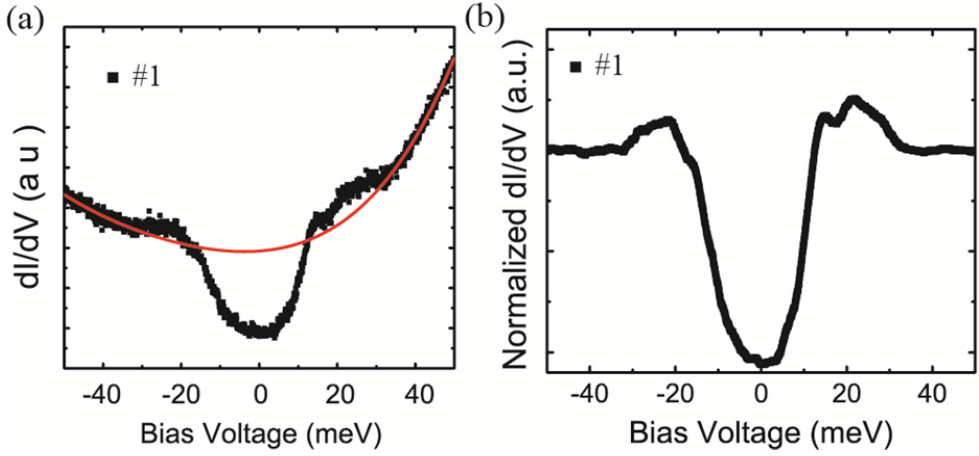

Fig. S1| (a) Original dI/dV spectrum of 1uc-FeSe/STO at point #1 marked in Fig. 3. The background for normalization is fitted by a polynomial function and plotted in the red line. (b) Normalized dI/dV spectra at point #1 as shown in Fig. 3(c). The original dI/dV spectra are normalized by the red line in (a) and smoothed by the average of adjacent 20 points.

## III. Parameters of different superconducting systems

Values of $k_F\xi_{pair}$, $\Delta/E_F$ and $T_C/T_F$ presented in the manuscript were determined based on the existing independent experimental measurements. The pair size $\xi_{pair}$ is roughly replaced by the phase coherent length $\xi_{phase}$, because in BCS regime the two length is only differed by a factor $3/\sqrt{2}$ [57]. The parameters of $k_F$, $\Delta$, $E_F$, and $T_C$ of 1uc-FeSe/STO were extracted from band structure measured by angle-resolved photoemission spectroscopy [33]. The parameters of other superconducting systems were extracted from references as listed in Table SI.

Table SI. Values of $k_F\xi_{pair}$, $\Delta/E_F$ and $T_C/T_F$ for different superconducting systems.

| Systems and References. | $k_F$ (1/Å) | $\xi_{pair}$ (Å) | $k_F\xi_{pair}$ | $\Delta$(meV) | $E_F$ (meV) | $\Delta/E_F$ | $T_C$ (K) | $T_F$ (K) | $T_C/T_F$ |
|---|---|---|---|---|---|---|---|---|---|
| FeSe bulk [17,18] | 0.065 | 50 | 3.25 | 2.5 | 2 - 3 | 0.83 - 1 | 8 | 23 - 35 | 0.229-0.348 |
| FeSe bulk [17,18] | 0.153 | 50 | 7.65 | 3.5 | 10.0 | 0.30 | 8 | 116 | 0.069 |
| (LiFe)OHFeSe [36,59] | 0.280 | 33 | 9.24 | 10.5 | 43.0 | 0.24 | 40 | 449 | 0.080 |
| $k_{0.8}$FeSe [37] | - | - | - | 10.3 | 60.0 | 0.17 | 32 | 696 | 0.046 |
| 1uc-FeSe/STO | 0.208 | 20[25] | 4.20 | 20.0 | 56.0 | 0.36 | 65 | 650 | 0.100 |
| 1uc FeSe/STO under-annealed [60] | - | - | - | 10.0 | 48.3 | 0.21 | 40 | 561 | 0.071 |
| | - | - | - | 14.0 | 50.6 | 0.28 | 55 | 587 | 0.094 |
| | - | - | - | 16.0 | 55.7 | 0.29 | 60 | 646 | 0.093 |
| | - | - | - | 19.0 | 54.9 | 0.35 | 65 | 637 | 0.102 |
| 1uc-FeSe/BTO [40] | - | - | - | 17.0 | 63.0 | 0.27 | 70 | 731 | 0.096 |
| | - | - | - | 19.5 | 63.0 | 0.31 | 75 | 731 | 0.103 |
| 1uc-FeSe/TiO$_2$ [38] | - | - | - | 14.6 | 45.0 | 0.32 | 63 | 522 | 0.121 |
| Cuprates 123 [11,13,15] | - | - | - | - | - | - | 54 | 1232 | 0.044 |
| | - | - | - | - | - | - | 62 | 1325 | 0.047 |
| | - | - | - | - | - | - | 69 | 1478 | 0.047 |
| | - | - | - | - | - | - | 75 | 1740 | 0.043 |
| | - | - | - | - | - | - | 90 | 1977 | 0.045 |
| | - | - | - | - | - | - | 87 | 2204 | 0.040 |
| | - | - | - | - | - | - | 79 | 2741 | 0.029 |
| | - | - | - | - | - | - | 91 | 2947 | 0.031 |
| | - | - | - | - | - | - | 91 | 3170 | 0.029 |
| Cuprates 2223 [11,13,15] | - | - | - | - | - | - | 110 | 2413 | 0.046 |
| | - | - | - | - | - | - | 126 | 3112 | 0.040 |
| | - | - | - | - | - | - | 108 | 3228 | 0.034 |
| Cuprates 214 [11,13,15] | - | - | - | - | - | - | 36 | 922 | 0.039 |
| | - | - | - | - | - | - | 30 | 812 | 0.036 |
| | - | - | - | - | - | - | 40 | 1561 | 0.026 |
| | - | - | - | - | - | - | 30 | 1709 | 0.018 |
| | - | - | - | - | - | - | 28 | 1709 | 0.016 |
| CeRhIn$_5$ [16] | - | - | - | - | - | - | - | - | 0.016 |
| Ce$_2$PdIn$_8$ [16] | - | - | - | - | - | - | - | - | 0.009 |
| UPt$_3$ [16] | - | - | - | - | - | - | - | - | 0.008 |
| Ce$_2$CoIn$_8$ [16] | - | - | - | - | - | - | - | - | 0.003 |
| NbSe$_2$ [11,14] | - | - | - | - | - | - | 7 | 802 | 0.009 |
| Nb [11,14] | - | - | - | - | - | - | - | 10 | 12607 | 7.62E$^{-04}$ |
| Sn [11,14] | - | - | - | - | - | - | - | 4 | 115638 | 3.26E$^{-05}$ |
| Zn [11,14] | - | - | - | - | - | - | - | 1 | 109496 | 7.98E$^{-06}$ |
| Al [11,14] | - | - | - | - | - | - | - | 1 | 133721 | 8.56E$^{-06}$ |